%% file: main.tex
\documentclass[utf8]{preprint}

\input{newcommands}
\def\Authors{Remy Gardier\,$^{1,*}$, Juan Luis Villarreal Haro\,$^{1}$, Erick J Canales-Rodríguez\,$^{1}$, Ileana O. Jelescu\,$^{2,3}$, Gabriel Girard\,$^{1,2, 4}$,  Jonathan Rafael-Patino\,$^{2,1}$, Jean-Philippe Thiran\,$^{1, 2, 4}$}

\def\Address{$^{1}$Signal Processing Laboratory (LTS5), École Polytechnique Fédérale de Lausanne (EPFL), Lausanne, Switzerland\\
$^{2}$Radiology Department, Centre Hospitalier Universitaire Vaudois (CHUV) and University of Lausanne (UNIL), Lausanne, Switzerland\\
$^{3}$School of Biology and Medicine, University of Lausanne (UNIL), Lausanne, Switzerland\\
$^{4}$Department of Computer Science, Universit\'e de Sherbrooke, Sherbrooke, Canada\\
$^{5}$CIBM Center for Biomedical Imaging, Switzerland}

\def\corrAuthor{Remy Gardier, Signal Processing Laboratory (LTS5), Station 11, CH-1015 Lausanne,
Switzerland}

\def\corrEmail{remy.gardier@epfl.ch}

\def\title{Cellular EXchange Imaging (CEXI): Evaluation of a diffusion model including water exchange in cells using numerical phantoms of permeable spheres}

\begin{document}
\onecolumn
    \sffamily
   \parindent0pt
   
    {\vspace{5cm}  \LARGE  \bfseries \title}
    
    {\vspace{0.5cm} \bfseries \Authors}
    
    {\vspace{0.5cm}  \scriptsize \itshape \Address{}} 
    
    {\vspace{0.5cm}   \footnotesize Correspondence*: \\ \corrAuthor \\ \corrEmail}

    {\vspace{0.5cm}  \textit{Accepted to Magnetic Resonance in Medicine - DOI: 10.1002/mrm.29720 } \\}

\begin{abstract}
\section{Purpose}Biophysical models of diffusion MRI have been developed to characterize microstructure in various tissues, but existing models are not suitable for tissue composed of permeable spherical cells. In this study we introduce Cellular Exchange Imaging (CEXI), a model tailored for permeable spherical cells, and compares its performance to a related Ball \& Sphere (BS) model that neglects permeability.
\section{Methods}  We generated DW-MRI signals using Monte-Carlo simulations with a PGSE sequence in numerical substrates made of spherical cells and their extracellular space for a range of membrane permeability. From these signals, the properties of the substrates were inferred using both BS and CEXI models.
\section{Results} CEXI outperformed the impermeable model by providing more stable estimates cell size and intracellular volume fraction that were diffusion time-independent. Notably, CEXI accurately estimated the exchange time for low to moderate permeability levels previously reported in other studies ($\kappa<25\mu m/s$). However, in highly permeable substrates ($\kappa=50\mu m/s$), the estimated parameters were less stable, particularly the diffusion coefficients.
\section{Conclusion} This study highlights the importance of modeling the exchange time to accurately quantify microstructure properties in permeable cellular substrates. Future studies should evaluate CEXI in clinical applications such as lymph nodes, investigate exchange time as a potential biomarker of tumor severity, and develop more appropriate tissue models that account for anisotropic diffusion and highly permeable membranes.

\tiny
 \keyFont{ \section{Keywords:} Diffusion MRI, Tumor microstructure, Exchange, Monte-Carlo Simulations, Compartmentalized model, Lymph node}
\end{abstract}

\section{Introduction}
Over the last decades, diffusion-weighted magnetic resonance imaging (DW-MRI) has been used to characterize tissue microstructure, particularly with biophysical models of healthy white matter \citep{Alexander2010, Zhang2012, Jespersen2007, Tariq2016, Kaden2016, Novikov2018, Novikov2019}, healthy gray matter \citep{Palombo2020a, Jelescu2022, Olesen2022} and tumors \citep{Reynaud2016, Karunanithy2019, Li2017, Jiang2017, Reynaud2016b}. These models group the biological entities into compartments with similar contributions to the diffusion signal and differ essentially by the number of geometric features, compartments and the targeted microstructure features. 

One of the challenges of biophysical models is their specificity to one particular tissue, which requires rethinking the optimal model and assumptions that best capture its features. For example, blood displacement inside the tumor vessels \citep{Panagiotaki2014, Ianus2020, Sahalan2018, Iima2014} induces anisotropic diffusion, which usually is not accounted for in the healthy white matter.

An important assumption of most biophysical models is negligible water exchange between compartments, which might not be valid in unmyelinated tissue \citep{Jelescu2022, Jelescu2020b, Yang2018}, urging for specific models that include the water exchange between the intracellular and extracellular compartments, for example to model gray matter \citep{Jelescu2022, Olesen2022} or cancerous tissue \citep{Zhao2008, Reynaud2017, Reynaud2016, Jiang2022}. The biophysical models that neglect the effect of water exchange on the signal employ acquisition sequences with short diffusion time \citep{Palombo2020a, Aggarwal2020, Reynaud2016, Panagiotaki2014}, while the ones that account for exchange either use non-PGSE gradient sequences \citep{Jiang2017, Jiang2022, Li2017, Lasic2011} or are based on the K\"arger model of exchange \citep{Karger1985, Stanisz1997, Jelescu2022, Olesen2022, Karunanithy2019}.

During the development of complex tissue models, their evaluation and validation in controlled environments helps identifying the best microstructure model and designing the optimal DW-MRI acquisition protocol. In this context, Monte-Carlo Diffusion Simulations (MCDS) \citep{Patino2020, Lee2021, Hall2009, Brusini2019} can be used to simulate the diffusion signals in complex substrates with known ground truth and without assuming an analytical equation, thus eliminating the comparison bias across different models. Moreover, realistic substrates \citep{Fieremans2010, Abdollahzadeh2019, Lee2020, Andersson2020, Patino2020} can be designed to study various microstructure parameters, including the membrane permeability \citep{Lee2021}, in a controlled manner.

In this work, the performances of compartment models excluding and including water exchange between compartments were studied employing MCDS in packed spheres with finite membrane permeability. More specifically, the work focused on the influence of non-negligible membrane permeability on microstructure estimation (i.e. the cell sizes, extracellular and intracellular diffusion coefficients and volume fractions) with a two-compartment model neglecting exchange terms (Impermeable Ball \& Sphere \citep{panagiotaki2012}) and a newly proposed Cellular Exchange Imaging (CEXI) model that accounts for exchange (Permeable Ball \& Sphere). 

\section{Methods}\label{sec:method_method}

Employing MCDS in numerical substrates with water exchange across cell membranes, we investigated the effects of permeability on microstructure estimation by compartment-based models. Section~\ref{sec:method_mcdc} details the simulation framework, the choice of the simulation parameters, and the numerical substrate. Next, Section~\ref{sec:method_imperm_to_perm} compares the diffusion regime of impermeable and permeable tissue. Finally, Section~\ref{sec:method_comp} describes the two-compartment models evaluated in this simulation framework.

\subsection{Monte-Carlo simulations}\label{sec:method_mcdc}
\subsubsection{Simulations in permeable substrates}
To simulate diffusion in permeable tissue \citep{Lee2021, Szafer1995}, we extended the MCDC Simulator \citep{Patino2020}. In non-exchanging media, the diffusion process inside different biological structures is assumed to contribute independently to the DW-MRI signal. Consequently, the intracellular and extracellular signals are usually generated individually and summed to produce the total signal. In permeable substrates, some particles cross the membrane, which requires implementing multiple-diffusivity features to generate intracellular and extracellular signals simultaneously (see supplementary materials for derivations). To minimize the boundary effects, we used a large voxel size and periodic boundary conditions when a particle crossed the voxel boundary \citep{Patino2020}. After simulating the particles' trajectories, we generated the DW-MRI signals with a Graphics Processing Unit (GPU) implementation for a PGSE sequence. 

\subsubsection{Lymph nodes imaging}\label{sec::histology}

The application that motivates this study is the discrimination of healthy from cancerous lymph nodes using non-invasive imaging. Several studies investigated the potential of dMRI to replace invasive procedures such as biopsy for lymph node metastases diagnosis, particularly the apparent diffusion coefficient (ADC) in breast~\citep{zaiton2016, zhao2020}, neck~\citep{suh2018}, and colorectal~\citep{yasui2009} cancer, or more advanced biophysical models for benign and malignant tumor discrimination in lymph nodes \citep{Ianus2020}.

Healthy lymph nodes (Figure~\ref{fig:histology_substrate}, left magnification) are homogeneous tissues composed of round lymphocytes that are known to be permeable \citep{Cheung1982, Grinstein1983, Wesselborg1993}. In the hematoxylin and eosin stain of Figure~\ref{fig:histology_substrate}, we measured using QuPath \citep{bankhead2017} a cell area of $56 \pm 24 \mu m^2$, corresponding to a radius from $3$ to $5\mu m$ in perfectly spherical cells, similar to previously reported values \citep{Ianus2020}. In tumors (Figure~\ref{fig:histology_substrate}, right magnification), the tissue is more heterogeneous with the presence of blood vessels and conjunctive tissue. We measured a larger mean cell area of $225 \pm 102 \mu m^2$, corresponding to cell radii from $6$ to 10 $\mu m$.

The present work focuses on membrane permeability in spherical cells to capture the effect of exchange on microstructure estimation. Our substrates are isotropic, but a previous study \citep{Ianus2020} showed that the biophysical model that accounts for anisotropic diffusion had better results in lymph node discrimination. However, we decided to focus on the effect of exchange, and future work will combine exchange and anisotropic diffusion to better capture the complexity of lymph nodes' microstructure.

\subsubsection{Numerical substrates design}\label{sec:method_mcdc_lymph}

In addition to the rules of the particles' dynamic, MCDS required a numerical substrate into which the particles diffuse. As motivated in Section~\ref{sec::histology}, we focus on numerical substrates made of packed spheres with finite membrane permeability.

We settled the substrates as isotropic voxels of side-length $100 \mu m$, filled with randomly placed spheres. These spheres had their radii normally distributed around a mean radius $R_s$ equal to  $\{2; 3; 4; 5; 8 \} \mu m$ with a standard deviation of 1\% of the radius. For all substrates, the intracellular volume fraction $\text{\textit{ICVF}}$ reached $0.65$, corresponding to the value reported in malignant lymph nodes \citep{Ianus2020}.

\begin{figure*}
\centerline{\includegraphics[width=\textwidth]{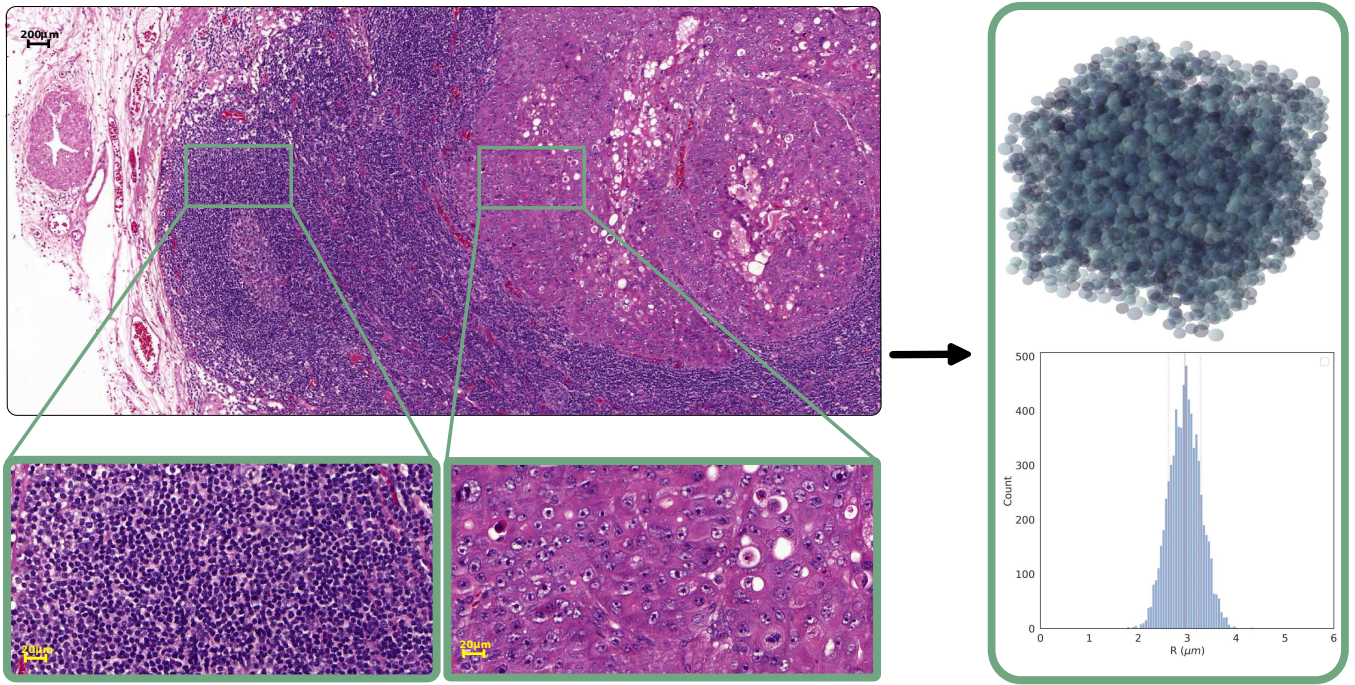}}
\caption{\ \textbf{Lymph node histology and numerical substrate.}  Example of lymph node histology and a numerical substrate (Table ~\ref{tab:experiments}) with a mean radius $R \simeq 3\mu m$. The histogram shows the radii distribution of the cells in the substrates. The solid red and dashed lines show the mean and the standard deviation of $R$, respectively. Histology was provided by Prof. Dr. phil. nat. Inti Zlobec. }\label{fig:histology_substrate}
\end{figure*}

Because the spheres in each substrate had different sizes, the apparent mean cell radius was weighted by the cell size distribution within the voxel. In the case of small cells, the apparent mean radius tends to $R_{small} = \left(\frac{< R^7 > }{<R^3>}\right)^{\frac{1}{4}}$ at long diffusion times while, in the narrow-pulse approximation, the apparent mean radius becomes $R_{NP} = \left(\frac{< R^5 > }{<R^3>}\right)^{\frac{1}{2}}$~\citep{Olesen2022}. Table~\ref{tab:experiments} summarizes the structural properties of the numerical substrates used in the experiments, and Figure~\ref{fig:histology_substrate} shows illustrative examples of the substrate $S_2$.

In addition to the structural properties, particles' trajectories depend on the biological properties of the substrates, which are the intracellular $D_{i, 0}$ and extracellular $D_{e, 0}$ diffusion coefficients and the membrane permeability $\kappa$. We chose their ranges in the simulations based on previously reported values for tumors \citep{Mukherjee2016} (Table~\ref{tab:experiments}).

\subsubsection{Sensitivity and reliability analysis of MCDS in permeable substrates}\label{sec:method_simulator_validation}
The first experiment aimed to evaluate the reliability and repeatability of the simulated signals with the substrates' biological properties (i.e. the permeability $\kappa$, the intracellular $D_{i,0}$ and the extracellular $D_{e, 0}$ diffusion coefficients) described in the Section~\ref{sec:method_mcdc_lymph}. We performed this sensitivity analysis with the substrate having a mean cell size of $2\mu m$ (Table~\ref{tab:experiments}, $S_1$), and we fixed the maximal particle density to $ \rho=2\mu m^{-3}$, i.e. two millions of particles. The membrane permeability $\kappa$ ranged from 0 to $ 50\mu m/s$, while we set the intracellular $D_{i, 0}$ and extracellular $D_{e, 0}$ diffusion coefficients to $1$ or $2 \mu m^2/ms$. For all combinations of these parameters, we calculated with bootstrapping the normalized mean square error (\textit{NMSE}s) of the DW-MRI signals generated with the PGSE sequence (Table~\ref{tab:experiments}) in $24$ uniformly distributed orientations on the unit sphere. 

\subsection{From impermeable to permeable tissue}\label{sec:method_imperm_to_perm}
\subsubsection{Structural disorder in impermeable substrates}\label{sec:method_structural_disorder}
The apparent intracellular/extracellular diffusivities and kurtosis in impermeable tissue have distinct time-dependency. In the intracellular space, the water molecules are confined in the cells, and the apparent intracellular diffusivity $ADC_{in}(t)$ tends to 0. Conversely, the water molecules of the extracellular space encounter obstacles but are free to diffuse. Therefore, the extracellular diffusivity $ADC_{ex}(t)$ converges to a non-zero long-time limit. Simultaneously, the kurtosis $AKC_{ex}(t)$ converges to zero, with a rate characterizing the mesoscopic disorder of the tissue \citep{novikov2014, Burcaw2015}. In our substrates, the structural-disorder theory predicts that the impermeable $ADC_{ex}(t)$ and $AKC_{ex}(t)$ converge following $\propto t^{-1}$ (3-dimensional diffusion within short-range disorder represented by random spheres). Therefore, we fitted the equations $ADC_{ex}(t) = ADC_{\infty} + A_D/t $ and $AKC_{ex}(t) = AKC_{\infty} + A_K/t $ to the $ADC_{ex}(t)$ and $AKC_{ex}(t)$ calculated from the diffusion propagator (estimated from the spins trajectories in the media) for three substrates (Table~\ref{tab:experiments} $S_1$, $S_3$, $S_5$) and two extracellular diffusion coefficients $D_{e,0} = \{1, 2\} \mu m^ 2/ms$. We chose those values according to simulation experiments of previous studies \citep{Palombo2020a, Olesen2022, Afzali2021, Karunanithy2019, Bonet2019}.

Because tracking the particles' relative position in the substrate is possible with MCDS, we independently calculated the intracellular and extracellular diffusion coefficients and kurtosis. To this end, we assigned a particle to a compartment at initialization for the entire simulation. In permeable substrates, the particles cross the cell membranes and therefore, the intracellular and extracellular water interact, especially for high permeability values and long diffusion times. In this regime, interpreting the model parameters, like the diffusivities, is more challenging since they are defined for water molecules residing inside the same compartment during the experimental time. Therefore, in the presence of exchange, we focused on the time-dependency of the $ADC(t)$ and the $AKC(t)$ (Section \ref{sec:method_karger}). 

\subsubsection{K\"arger assumptions validity in permeable substrates}\label{sec:method_karger}
The K\"arger model of exchange is valid under two assumptions \citep{Karger1985}, under which water diffuses in the barrier-limited regime. First, the diffusion inside each compartment is Gaussian, meaning the diffusion time is long enough for the particles to explore the compartment. 
This implies a time-independent diffusion coefficient. The second assumption links these characteristic times to membrane permeability by requiring slow exchange between the compartments, i.e. the exchange time must be longer than the characteristic times of the compartments.

We determined the range of validity of the K\"arger model using a time-dependency analysis of the $ADC(t)$ and the $AKC(t)$ of the same substrates as presented in Section~\ref{sec:method_structural_disorder} (Table~\ref{tab:experiments}, $S_1$, $S_3$, $S_5$) for different permeability values ($\kappa = \{0, 10, 25, 50\} \mu m / s$), spanning the parameter range considered in this work. From their time-dependencies, we quantified the effect of permeability on the estimated $ADC(t)$ and $AKC(t)$ for each experiment. Ultimately, we used these observations to discuss the validity of the model's assumptions (Section~\ref{sec:discussion_model}). 

\subsubsection{Diffusion regime in the presence of exchange}\label{sec:method_diffusion_regime}
In impermeable tissue, the tissue structure governs the diffusion dynamics. In the presence of exchange, the tissue structure and the exchange compete, producing different effects on the signal \citep{Olesen2022}. At a fixed diffusion encoding, i.e. $b$-value, the evolution of the signal with the diffusion time is different whether the water exchange through the membrane or the restriction due to structure dominates. If the diffusion is barrier-limited, the K\"arger model predicts a decreasing signal with the diffusion time~\citep{Karger1985, Fieremans2010}. Conversely, the signal increases if the restriction dictates the diffusion dynamics~\citep{Neuman1974, Palombo2020a}.

To determine which effect dictates diffusion in our experiments, we generated DW-MRI signals with the PGSE sequence of Table~\ref{tab:experiments} for the same parameters of the previous experiment (Section \ref{sec:method_karger}).

\subsection{Compartmentalized models of microstructure}\label{sec:method_comp}
From Section~\ref{sec:method_imperm_to_perm}, we identified the different diffusion regimes in our set of parameters. In this section, we describe two models of tumor microstructure: one designed for impermeable cellular tissue (Section~\ref{sec:ball}) and one for permeable tissue (Section~\ref{sec:cexi}).
\begin{equation}
         \fontsize{8}{10}\selectfont
        D_i(\Delta, \delta, R_s) = \frac{2}{ \delta^2 D_{i, s} \left(\Delta - \frac{\delta}{3}\right)} \left[\sum_{m=1}^{\infty} \frac{\alpha_m^{-4}}{\alpha_m^2R_s^2 - 2} \left[ 2\delta - \frac{2+e^{-\alpha_m^2 D_{i,s} \left(\Delta-\delta\right)} -2e^{-\alpha_m^2D_{i,s}\delta} - 2e^{-\alpha_m^2D_{i,s}\Delta} + e^{-\alpha_m^2D_{i,s} \left(\Delta+\delta\right)}} {\alpha_m^2D_{i,s}} \right]\right],
        \label{eq:Di_neumann}
\end{equation}

\subsubsection{Ball \& Sphere diffusion: An impermeable model of tumors}\label{sec:ball}
Under the assumption of impermeable membranes, the extracellular and intracellular compartments were modeled independently. Diffusion inside the compartments was assumed to follow analytical equations that depended on their geometry \citep{panagiotaki2012}. We modeled the intracellular compartment as an impermeable sphere of radius $R_s$ and diffusivity $D_{i,s}$ \citep{Neuman1974} (Equation~\ref{eq:Di_neumann}), where $\alpha_m$ is the $m^{th}$ root of $\left(\alpha R_s\right) J_{\frac{3}{2}}\left(s\alpha R_s\right) =  J_{\frac{5}{2}}\left(\alpha R_s\right)$ with $J_n(x)$ the Bessel function of the first kind, and extracellular diffusion with Gaussian isotropic diffusion of diffusivity $D_{e}$. Their respective volume fractions weighted the contribution of each compartment to the signal: $S_{BALL} = (1-f_i) e^{-b D_e} + f_i e^{-b  D_i}$. This model had four parameters: the diffusion coefficients $D_{i,s}$, $D_e$, the cell size $R_s$ and the intracellular volume fraction $f_i$.

\subsubsection{Cellular EXchange Imaging: A permeable model for spherical tumors}\label{sec:cexi}
The model described in Section~\ref{sec:ball} is valid in impermeable substrates only. Conversely, our Cellular EXchange Imaging (CEXI) model is a two-compartment model that includes exchange based on the K\"arger model \citep{Karger1985} between an intracellular compartment that models spherical cells and an extracellular space. Similarly to the impermeable model, we modeled the spherical cells of radius $R_s$ and intra-diffusivity $D_{i,s}$ by a spherical compartment with an apparent diffusion coefficient $D_i$ given by Eq.~\ref{eq:Di_neumann}. 

We assumed Gaussian and isotropic diffusion in the extracellular space with a diffusivity $D_{ex}$. The water exchange rates from the intracellular to extracellular compartments $k_i$ and from extracellular to intracellular $k_{ex}$ satisfied the spin conservation relation $k_i f_i = k_{ex}(1-f_i)$, where $f_i$ is the intracellular volume fraction. In spherical cells, the exchange rate $k_i$ and the membrane permeability $\kappa$ are linked by the relation $k_{i} = (1-f_i) (3\kappa / R_s)$ \citep{Fieremans2010}. Within this framework, the magnetization of the extracellular $M_{ex}$ and the intracellular $M_i$ compartments follow the differential equations  \citep{Stanisz1997}
\begin{equation}
    \left\{
    \begin{aligned}
    \frac{d M_{ex}}{dt} & = - (q^2 D_e + k_{ex}) M_{ex} + k_i M_i,\\
    \frac{d M_i}{dt} & = k_{ex} M_{ex} - (q^ 2 D_i + k_i) M_i,    
    \end{aligned} 
    \right.
    \label{eq:CEXI_ODE}
\end{equation}

with the initial conditions $M_{ex}\rvert_{t=0} = (1-f_i)$ and $M_{i}\rvert_{t=0} = f_i$,  and where $q^2$ was the wavenumber of the PGSE sequence. Ultimately, the signal was the weighted sum of these magnetizations $S_{CEXI} = (1-f_i) M_{ex} + f_i M_i$. With this formulation, CEXI had five parameters: the intracellular $D_{i,s}$ and extracellular $D_{e}$ diffusion coefficients, the membrane permeability $\kappa$, the intracellular signal fraction $f_i$ and the cell radius $R_s$. 

Under K\"arger assumptions, Eqs.\ref{eq:CEXI_ODE} have an analytical solution \citep{Jelescu2022, Karunanithy2019}. However, the time-dependency of the intracellular diffusivity in large cells might invalidate this model. For this reason, we integrated Eqs.\ref{eq:CEXI_ODE} for each diffusion time separately with an intracellular diffusion coefficient $D_i$ calculated with Eq.\ref{eq:Di_neumann} for each particular diffusion time, and one extracellular diffusion coefficient $D_e$ common to all signals. This approach allowed us to consider the time dependency of $D_i$ empirically. We acknowledge that the K\"arger formulation is technically invalid in the case of time-dependent diffusion in one compartment, and our approach constituted an approximation similarly to previous studies \citep{Jelescu2022, Fieremans2010, Karunanithy2019}. This approximation was supported by the previous experiments (Section \ref{sec:result_disorder}, Figure~\ref{fig:ADCin_ADCex_imperm}), where we showed that the $ADC_{ex}(t)$ in impermeable tissue converged quickly to its long-time limit to a value nearly independent of the cell size, conversely to the intracellular diffusivity, which converged more slowly to its long-time limit.

\subsubsection{Performance in permeable tissue}\label{sec:method_model}
In this last experiment, we compared the performance of the impermeable Ball \& Sphere model to the CEXI model. Because the numerical substrates of this work were made of spheres only, we did not include three-compartment models such as VERDICT \citep{Panagiotaki2014} nor gray matter models that account for neurites \citep{Palombo2020a, Olesen2022, Jelescu2022}.

From the simulations described in Section~\ref{sec:method_mcdc}, we selected the subset having an extracellular diffusivity of $ 2 \mu m^2 /ms$. We generated the DW-MRI signals with the PGSE sequence described in Table \ref{tab:experiments}. From the noise-free signals, we generated using the DIPY software library \citep{Garyfallidis2014} 30 corrupted signals with Rician noise with a signal-to-noise (\textit{SNR}) ratio of 30 and 80.

We fitted the models for all $\Delta$ simultaneously and accounted for the noise~\cite{Alexander2008} with the constrained least squares implementation of the python optimization library LEVMAR \citep{lourakis04LM}. We constrained the signal fractions to sum up to 1. We imposed boundaries on the value of the parameters to avoid unrealistic estimations: $\overline{R} \in \left[0.1, 20\right] \mu m$, $\overline{\text{\textit{ICVF}}} \in \left[0.1, 0.9\right]$,  $(\overline{D}_i,\ \overline{D}_e)  \in \left[0.01, 3\right] \mu m^2/ms$, and $\kappa  \in \left[0, \infty\right[\mu m/s$. We performed ten optimizations with random initialization and selected the estimation that minimized the cost function.

We compared the models based on their estimates of the mean cell radius $\overline{R}$, intracellular volume fraction $\overline{\text{\textit{ICVF}}}$ and extracellular $\overline{D}_e$ and intracellular $\overline{D}_i$ diffusion coefficients. We calculated the ground truth values of the intracellular volume fraction $\text{\textit{ICVF}}$ and the volume-weighted mean cell radius $R$ from the substrates (Table~\ref{tab:experiments}). In Section~\ref{sec:method_imperm_to_perm}, we showed that the apparent extracellular diffusion coefficient $ADC_{ex}(\Delta)$ deviated significantly from $D_{e, 0}$ due to the obstacles encountered by the particles. Therefore, we considered as ground truth the $D_e$ calculated from the propagator of the extracellular compartment at the longest diffusion time in the impermeable substrates (i.e., $D_e = ADC_{ex}(t=\Delta_{max}, \kappa=0)$). 

\section{Results}\label{sec::results}
\subsection{Influence of the substrate properties and the simulation parameters on the Monte-Carlo simulated DW-MRI signals}
\begin{figure*}
\centerline{\includegraphics[width=\textwidth]{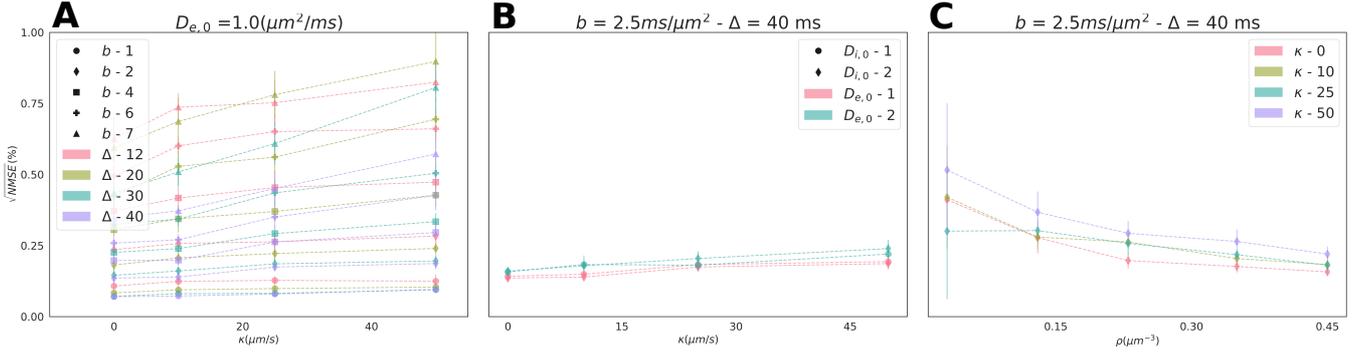}}
\caption{\ \textbf{Normalised mean squared error on the signal.} Normalised mean squared error (\textit{NMSE}) on the signal with the permeability $\kappa$ from bootstrapping for all pairs of $b$ and diffusion time $\Delta$ (A), for all diffusion coefficients $\Delta$ (B) and different particle density (C). (A) The \textit{NMSE} is shown for $D_{e, 0}=1 \mu m^2/ms$ and $D_{i, 0}=2\mu m^2/ms$. The error increased with $\kappa$ for all $b$ - $\Delta$ pairs, and its range for different $\Delta$ broadened as $b$ increased. (B) The \textit{NMSE} is shown for the pair ($b=2.5 ms\mu / m^2, \Delta=40ms)$, for different $D_{e, 0}$ (color) and  $D_{i, 0}$ (symbol). The error increased with $D_{e,0}$ but seemed independent on $D_{i,0}$. (C) The NMSE is plotted against the particle density used in the simulations for different permeability levels (color).  Simulations with a faster permeability had a larger NMSE for all particle densities. The $\sqrt{\text{\textit{NMSE}}}$ with the chosen simulation parameters never exceeded $1\%$.}\label{fig:signal_error}
\end{figure*}

Figure~\ref{fig:signal_error} shows the evolution of the bootstrapped \textit{NMSE} on the DW-MRI signals. For the given diffusion coefficients (Figure~\ref{fig:signal_error}A), the \textit{NMSE} increased with $b$ and, at large $b$ ($b > 4 ms/\mu m^2$), the error became also dependent on $\Delta$ and $\kappa$. At fixed $(b=2.5 ms/\mu m^2, \Delta =40ms)$ (Figure \ref{fig:signal_error}B), the diffusion coefficients had a distinct effect on the error. Indeed, the \textit{NMSE} was dependent on $D_{e,0}$ (colors) while it seemed independent on $D_{i, 0}$ (symbols). Figure~\ref{fig:signal_error}C shows the decreasing rate of the \textit{NMSE} with the particle density used in the simulations. Even with a particle density of $0.5\mu m^{-3}$, the maximal error reached with the largest $b$, the longest $\Delta$ and the most permeable membrane remained under $1\%$ with the chosen simulation parameters. 

\subsection{Mesoscopic disorder: Time-dependency of $ADC_{ex}(t)$ and the $AKC_{ex}(t)$ in impermeable tissue}\label{sec:result_disorder}
\begin{figure*}
\centerline{\includegraphics[width=\textwidth]{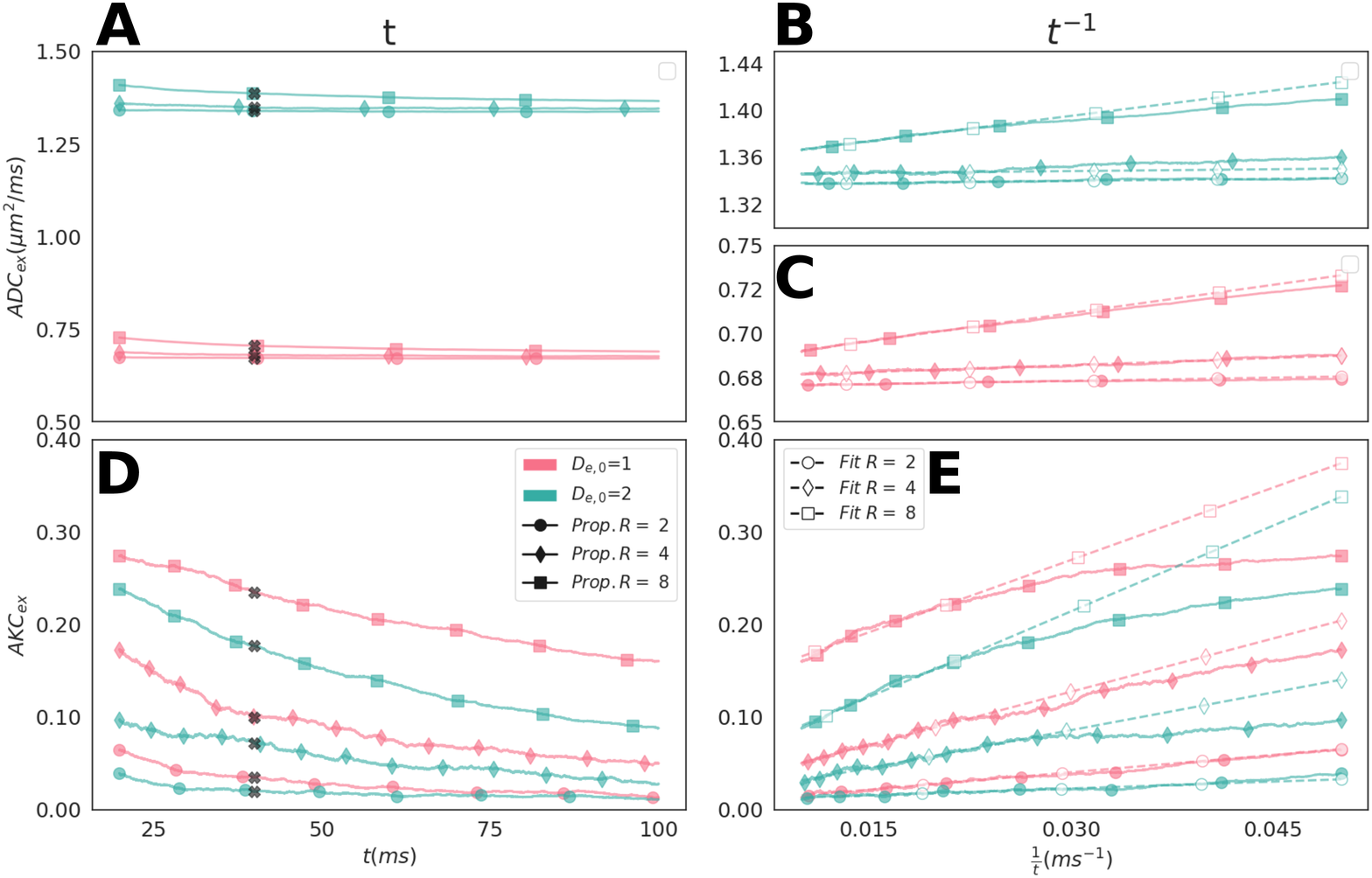}}
\caption{\ \textbf{ Extracellular $\mathbf{ADC_{ex}(t)}$ and $\mathbf{AKC_{ex}(t)}$ time-dependency in impermeable substrates.} Time-dependency of the $ADC_{ex}(t)$ (A, B, C) and the $AKC_{ex}(t)$ (D, E) of the extracellular compartment of the substrate $S_1$ (circle), $S_3$ (diamond) and $S_5$ (square) (Table~\ref{tab:experiments}) without exchange calculated from the propagator, for different extracellular diffusion coefficient (color). (A) The $ADC_{ex}$ was dependent on the extracellular diffusion coefficient but independent of the cell size. (D) Conversely, the time-dependency of the $AKC_{ex}$ strongly depended on the extracellular diffusion coefficient and the cell size.  (B, C) and E show the $ADC_{ex}(t)$ and $AKC_{ex}(t)$ against $t^{-1}$ as predicted by the structural-disorder theory~\citep{novikov2014, Burcaw2015}, respectively. The white symbols show the fit of the corresponding function to the data. Because $t^{-1}$ is decaying function of $t$, we fitted the equation in the decaying phase of the $ADC_{ex}(t)$ and $AKC_{ex}(t)$. The black crosses in A and D show the first point of the fitting.}
\label{fig:ADC_AKC_disorder}
\end{figure*}

Figure~\ref{fig:ADC_AKC_disorder} shows the time-dependency of the extra-cellular diffusion coefficient $ADC_{ex}(t)$ and kurtosis $AKC_{ex}(t)$ of the impermeable substrates $S_1$, $S_3$ and $S_5$, and the fitted parameters are summarized in Table~\ref{tab:mesoscopic_fit}. The rate of convergence of the $ADC_{ex}(t)$ depended on the cell size (symbols). In contrast, the long-time limits ($ADC_{\infty}$ and $AKC_{\infty}$ in Table.\ref{tab:mesoscopic_fit}) depended on the genuine extra-cellular diffusion coefficient $D_{e,0}$ (colors)(Figure \ref{fig:ADC_AKC_disorder}A and Table~\ref{tab:mesoscopic_fit}, $D_{\infty}$). Conversely, the $AKC_{ex}(t)$ showed a convergence rate that depended more on $D_{e,0}$ than the cell size (Figure \ref{fig:ADC_AKC_disorder}D).

Figure \ref{fig:ADC_AKC_disorder}B, C, and D show the $ADC_{ex}(t)$ and $AKC_{ex}(t)$ against $t^{-1}$, respectively, and the \textit{MSE} is reported in Table \ref{tab:mesoscopic_fit}. 

\subsection{K\"arger model: Time-dependency of $ADC(t)$ and the $AKC(t)$ in permeable tissue}
\begin{figure*}
\centerline{\includegraphics[width=\textwidth]{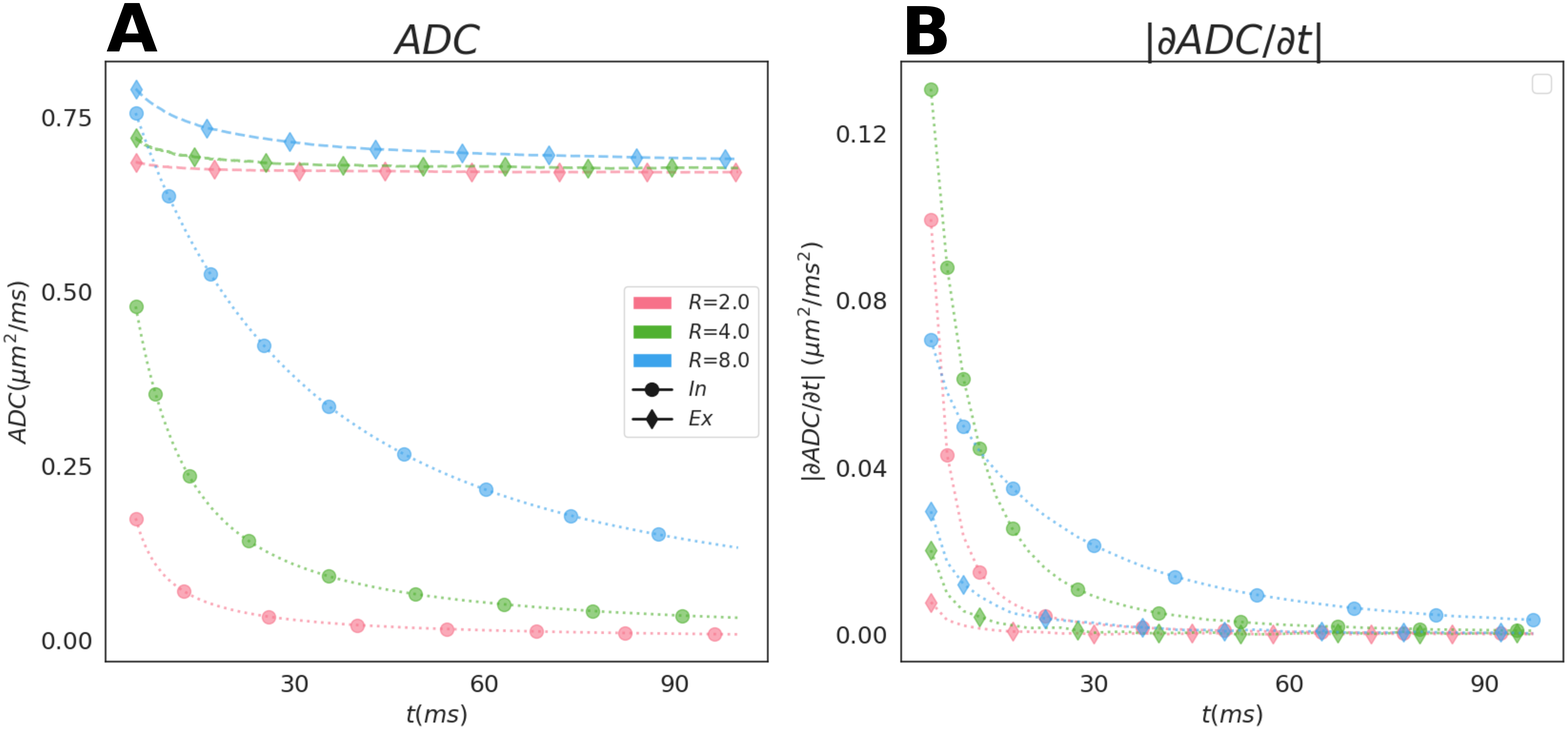}}
\caption{\ \textbf{ Extracellular $\mathbf{ADC_{ex}(t)}$ and intracellular $\mathbf{ADC_{in}(t)}$ time-dependency in impermeable substrates.} (A) Time-dependency and (B) derivatives of the extracellular $ADC_{ex}(t)$ (diamond) and intracellular $ADC_{in}(t)$ (circle) of the substrate $S_1$ (red), $S_3$ (green) and $S_5$ (blue) (Table~\ref{tab:experiments}) without exchange calculated from the propagator. The intracellular $D_{i, 0}$ and extracellular $D_{e, 0}$ diffusion coefficient were equal to $1 \mu m^2/ms$. (A) The $ADC_{ex}(t)$ converged faster than the $ADC_{in}(t)$ to its long-time limit. (B) Consequently, the rate of change $\partial ADC_{ex}/\partial t$ approached zero at short diffusion time, while the $\partial ADC_{in}/\partial t$ might not be negligible in our range of diffusion time.} \label{fig:ADCin_ADCex_imperm}
\end{figure*}

Figure~\ref{fig:ADCin_ADCex_imperm} shows the evolution of extracellular $ADC_{ex}(t)$ and intracellular $ADC_{in}(t)$ diffusion coefficients of the impermeable substrates $S_1$, $S_3$ and $S_5$ along each other (A) and their derivatives (B). As pointed out in Section \ref{sec:result_disorder}, the $ADC_{ex}(t)$'s converged quickly to the same long-time limit. Conversely, the convergence rate of the intracellular $ADC_{in}(t)$'s depended on the cell size of the substrates. The difference in decaying rate was confirmed by a large derivative of the intracellular diffusion coefficient for a longer diffusion time, in comparison to the extracellular $ADC_{ex}(t)$ (Figure~\ref{fig:ADCin_ADCex_imperm}B).

\begin{figure*}
\centerline{\includegraphics[width=\textwidth]{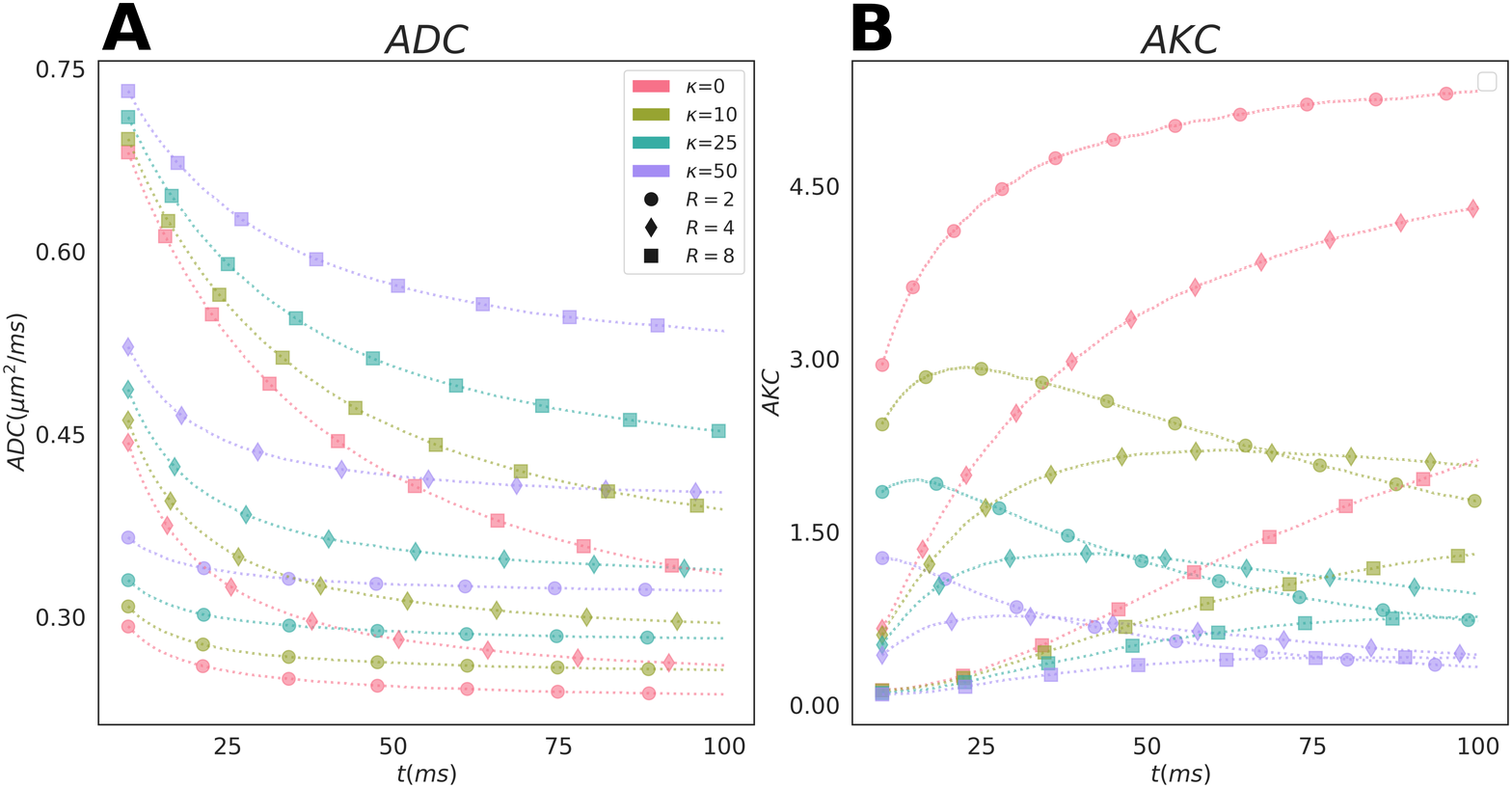}}
\caption{\ \textbf{$\mathbf{ADC(t)}$ and $\mathbf{ADK(t)}$ time-dependency in permeable substrates.} Time-dependency of the $ADC(t)$ (A) and $AKC(t)$ (B) in the three substrates  $S_1$ (circle), $S_3$(diamond) and $S_5$ (square) (Table~\ref{tab:experiments}) for an increasing permeability (color). The intracellular $D_{i, 0}$ and extracellular $D_{e, 0}$ diffusion coefficient were equal to $1 \mu m^2/ms$. (A) As the permeability increased, the $ADC(t)$ converged faster to its long-time limit. (B) Simultaneously, the $AKC(t)$ reached its maximum value and, therefore, its decaying regime for a shorter diffusion time, except for the impermeable substrates and the permeable cells with the biggest radii where $AKC(t)$ always grew. 
}\label{fig:ADC_AKC_perm}
\end{figure*}

Figure~\ref{fig:ADC_AKC_perm} shows the results of the time-dependency analysis of the $ADC(t)$ and the $AKC(t)$ in impermeable and permeable substrates. In larger cells (symbols), the $ADC(t)$ converged to its long-time limit slowly (Figure~\ref{fig:ADC_AKC_perm}A). This long-time limit of the $ADC(t)$ increased for an increasing permeability (colors). 

In impermeable substrates, the $AKC(t)$ increased over the range of simulated time (Figure~\ref{fig:ADC_AKC_perm}B). As soon as the membranes were permeable, the $AKC(t)$ exhibited a time dependency that could be split into two phases. The $AKC(t)$ increased to a peak value before decreasing to its long-time limit. This peak's location and intensity depended on the cell size and the membrane's permeability. 

\subsection{Diffusion regime: Barrier-limited or structure-limited?}
\begin{figure*}
\centerline{\includegraphics[width=\textwidth]{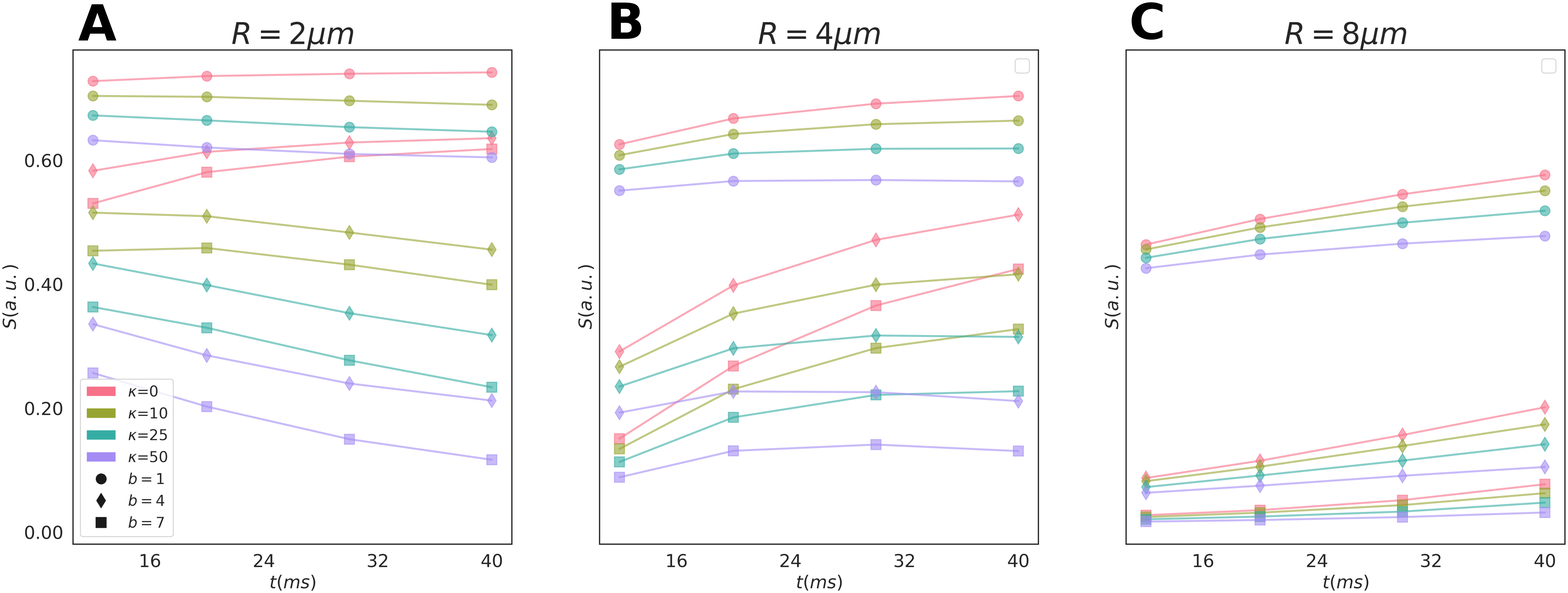}}
\caption{\ \textbf{Evolution of the DW-MRI signal with the diffusion time at fixed $\mathbf{b}$-value in permeable substrates.} The plots show the evolution of the DW-MRI signal with the diffusion time at fixed $b$-value (symbols) for different permeability (color) in the substrates $S_1$ (A), $S_3$ (B) and $S_5$(C) (Table~\ref{tab:experiments}). (A) In substrate $S_1$, the signal decayed with the diffusion time for all $b$ if the membranes were permeable. (C) Conversely, the signal increased for all $b$ and permeability in the substrate $S_5$. (B) In the substrate $S_3$ having an intermediate cell size, the signal increased at slow permeability ($\kappa \leq 10 \mu m/s$) but reached a plateau and started decreasing at faster permeability.}\label{fig:sig_bvals_perm}
\end{figure*}

Figure~\ref{fig:sig_bvals_perm} shows the evolution of the signal with the diffusion time at a fixed $b$ value. In the case of impermeable membranes (red), the signal increased with the diffusion time for all substrates and $b$ values (symbols), as expected from restriction. Each substrate had a different diffusion regime as soon as the membranes were permeable. In the smallest spheres $S_1$ (Figure~\ref{fig:sig_bvals_perm}A), the signals of all $b$ decreased with the diffusion time for all non-zero permeability values $\kappa$ (colors), an exchange-dominated regime. In contrast, the signals of the largest spheres $S_5$ (Figure~\ref{fig:sig_bvals_perm}C) increased for all $b$ and $\kappa$, a restriction-dominated regime. Finally, the signal of the substrate $S_3$ exhibited an intermediate regime (Figure~\ref{fig:sig_bvals_perm}B). At moderate permeability $\kappa < 25 \mu m/s$, the signal increased with the diffusion time, while at the largest permeability, it increased at a short diffusion time before decreasing (Figure~\ref{fig:sig_bvals_perm} blue, purple lines).  

\subsection{Microstructure model estimates in permeable tissue}\label{sec:results_model}
\begin{figure*}
\centerline{\includegraphics[width=\textwidth]{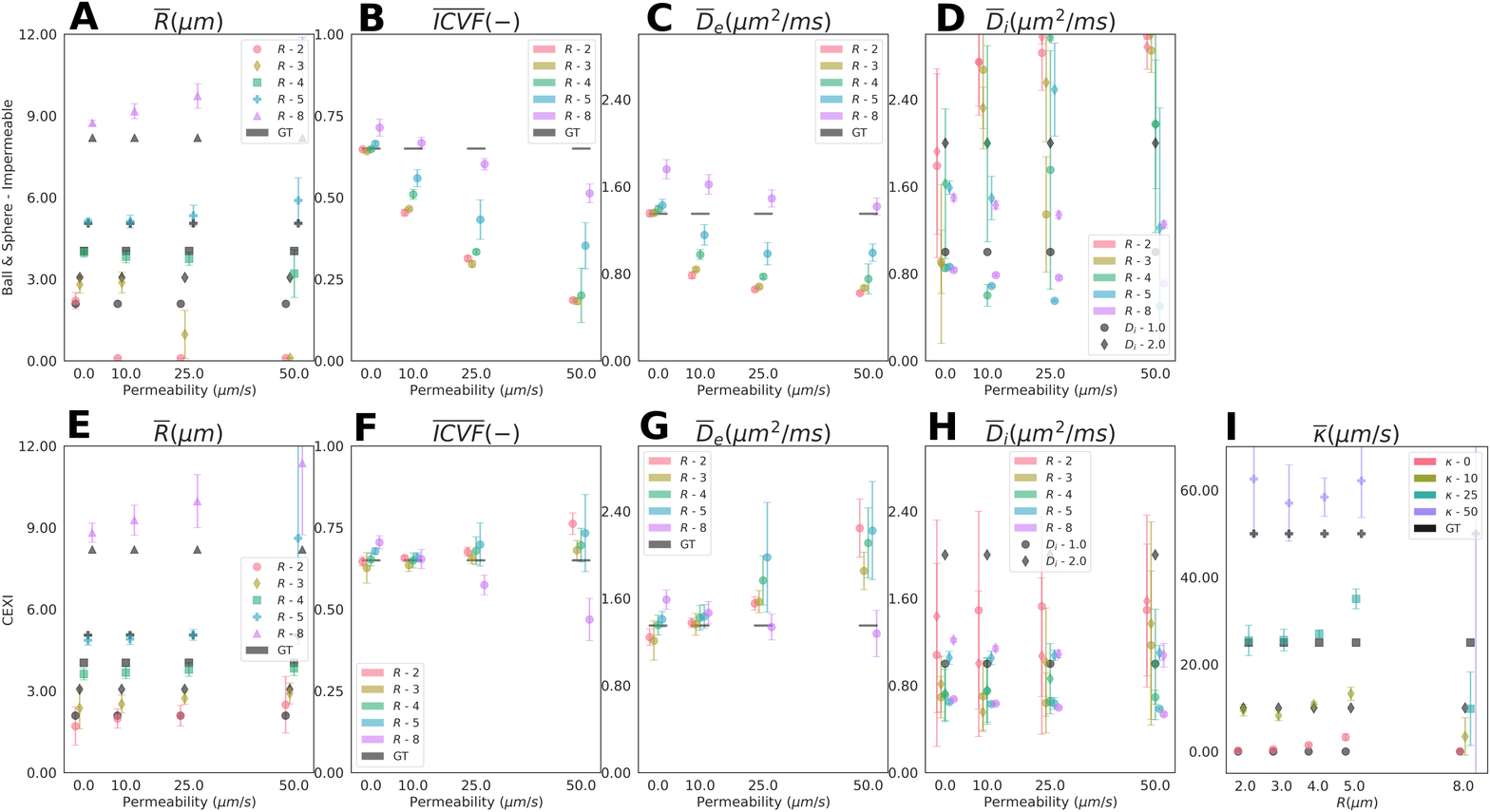}}
\caption{\ \textbf{Model estimates}. The plots show the estimates of the impermeable Ball \& Sphere (A-D) and CEXI (E-I) models from 30 DW-MRI signals with an \textit{SNR} of 80. The estimates of the cell size (A, E), the intracellular volume fraction (B, F), the extracellular (C, G) and intracellular (D, H) diffusion coefficients are plotted against the permeability. The CEXI permeability estimate is plotted against the cell size (I). The color encodes the cell size (A-H) or the permeability (I), and the black lines show the ground truths of the substrate parameters. The symbols and the bars show the estimates' mean and variance, respectively. }\label{fig:model_estimates}
\end{figure*}

Figure~\ref{fig:model_estimates} shows the substrate cell sizes $R$ (A, E), intracellular volume fractions $\text{\textit{ICVF}}$ (B, F), extracellular $D_e$ (C, G) and intracellular $D_i$ (D, H) diffusion coefficients estimated by the impermeable Ball \& Sphere and CEXI models, and the CEXI estimate of the permeability (I). The markers and the bars are the means and variances of the estimates, respectively. 

On the one hand, the impermeable Ball \& Sphere model (Figure~\ref{fig:model_estimates}A-D) estimated the cell size, intracellular volume fraction and extracellular diffusion coefficient of all impermeable substrates accurately, except for the largest spheres (purple). At larger permeability, the impermeable Ball \& Sphere model overestimated the cell size and underestimated the intracellular volume fraction more. Also, the estimated extracellular diffusivity decreased and stabilized at a smaller value.  Finally, the impermeable Ball \& Sphere model could not consistently estimate the intracellular diffusivity. 

On the other hand, the CEXI model (Figure~\ref{fig:model_estimates}D-H) estimates of the cell size were consistent for all permeability values. CEXI slightly underestimated the small cell sizes (Figure~\ref{fig:model_estimates}, red, yellow, green) and overestimated the largest cell size (purple). Similarly, CEXI provided stable estimates of the intracellular volume fraction and extracellular diffusion coefficient in the small cell sizes from low to moderate permeability levels ($\kappa \leq 25 \mu m /s $). CEXI tended to overestimate the true extracellular diffusion coefficient at increasing permeability, and the results were not accurate for permeability $\kappa \geq 25 \mu m / s$. Finally, CEXI provided stable estimates of the intracellular diffusivity $D_i$ in the substrates with a cell size $R>3 \mu m$.

In addition to these four parameters, CEXI estimated the exchange time $\tau_{ex}$ and, therefore, the permeability $\kappa$ between the intracellular and extracellular compartments (Figure~\ref{fig:model_estimates}I). For all ground truth permeability values (color), the mean and the variance of the estimated permeability $\overline{\kappa}$ increased with the cell size and the true permeability. At moderate permeability $\kappa \leq 25 \mu m/s$ and in the small cell size $R < 4 \mu m$, CEXI estimated the permeability accurately. For all parameters, a lower \textit{SNR} led to a larger variance but comparable results (See Figures~\ref{fig:model_estimates} and Figure 1 of the supplementary materials for the results with an \textit{SNR} of 80 and 30, respectively).

\section{Discussion}\label{sec::discussion}
\subsection{Validation of the MCDS's in permeable tissue}\label{sec:discussion_signal}
This work investigated through realistic MCDS's the effect of membrane permeability $\kappa$ on the estimation of microstructure model parameters in a model of permeable spheres. To perform simulations in permeable substrates, we extended the open-source MCDC diffusion simulator \citep{Patino2020} following previous studies \citep{Szafer1995, Lee2021}.  

\subsubsection{Repeatability of the signal generation}
The permeability $\kappa$ was identified as the first important substrate parameter that determined the choice of simulation parameters, to guarantee the repeatability of the simulation and a small \textit{NMSE}, especially at a high $b$-value. The increase in the \textit{NMSE} with $\kappa$ was faster at high $b$, and the \textit{NMSE} range for different diffusion times $\Delta$ broadened as $b$ increased. This suggests that the particle density required for the simulations was more dependent on $b$ and $\kappa$ than $\Delta$. In parallel, the extracellular diffusion coefficient $D_{e, 0}$ was shown to play a major role in signal generation, which confirmed that both intracellular and extracellular signals must be simulated simultaneously. Despite the increase in the \textit{NMSE} with $\kappa$ and $b$, the small error $(\sqrt{\text{\textit{NMSE}}} < 1\%)$ of the signals validated the choice of the simulation parameters (Table~\ref{tab:experiments}) of the experiments.

\subsubsection{Is the simulated signal consistent?}
Simultaneously, a time-dependency analysis of the apparent diffusivity and kurtosis in impermeable substrates validated the correct implementation of the simulations by comparing to theoretically expected trends.

In the intracellular space, the $ADC_{in}(t)$ of the particles confined inside the cells converged to 0 at a rate dependent on the cell size. In the extracellular space, the diffusion coefficient $ADC_{ex}(t)$ converged quickly to its long-time limit both in small and large spheres. Interestingly, this long-time limit depended on the packing density of the substrate rather than the cell size. The convergence rate of the $ADC_{ex}(t)$ and the kurtosis $ADK_{ex}(t)$ were consistent with the power-law of $t^{-1}$ predicted by structural disorder in a 3D environment of spheres (short-range 3D disorder) \citep{novikov2014, Burcaw2015} in the long time limit.
 
\subsubsection{Diffusion regime in tissues of different permeability and cell sizes}\label{sec:discussion_diffusion_regime}
In permeable substrates, the physical interpretation of the $ADC_{in}(t)$ and the $ADC_{ex}(t)$ became ambiguous, and we focused our attention on the total $ADC(t)$ and $AKC(t)$. At larger permeability values, the mixing rate of the compartments increased, and the $ADC(t)$ converged faster. Similarly, the time dependency of the $AKC(t)$ depended on the permeability. As the permeability increased, the peaking time of the $AKC(t)$ shifted to shorter diffusion times, coherently with previous work \citep{Aggarwal2020, Zhang2021}. 

Permeability added another degree of freedom to the diffusion regime, which modulated the DW-MRI signal. Following recent work~\cite{Olesen2022}, we showed that the effects of exchange and restriction competed in the signal dynamic. For the cell sizes investigated in this work, the extremes had a clear but distinct diffusion regime. The water exchange dominated the diffusion dynamic in small spheres, while the restriction dominated in substrates with large spheres. Interestingly, the intermediate cell size exhibited a shift of diffusion dynamic at higher permeability between exchange and structure. The permeability at which this shift occurred ($\kappa=25\mu m / s$) corresponds to the minimal permeability value for which the $AKC(t)$ reached its peak in the range of diffusion time of the PGSE sequence, independently of the $b$-value. In other words, the diffusion time at which the diffusion regime shifted from restriction-dominated to exchange-dominated corresponded to the maximal value of the kurtosis $AKC(t)$. Therefore, it should be possible to determine the minimal diffusion time to enter the exchange-dominated diffusion regime from kurtosis measurements at multiple diffusion times. This provides a way to choose the parameters of the PGSE sequence in experiments, especially the minimal diffusion time, to become sensitive to the  microstructure feature of most interest. However, we noted that, experimentally, the range of achievable diffusion times might also depend on hardware limitations.

\subsection{Compartmentalized models in permeable substrates of tumors}\label{sec:discussion_model}
The last experiment compared the impermeable Ball \& Sphere model to our CEXI model, which included the water exchange. Previous studies demonstrated that estimating the substrate's properties with compartmentalized models is a challenging ill-posed problem. The low sensitivity of these models to the compartment diffusivities was highlighted in white matter \citep{Jelescu2016, Li2017} and, more recently in gray matter \citep{Palombo2020a, Jelescu2022, Olesen2022}. Additionally, the size of the axons in white matter \citep{Burcaw2015} or the cells in gray matter \citep{Afzali2021, Palombo2021, Olesen2022} was shown to be overestimated. This experiment showed the impact of permeability on the estimated model parameters, including the mean cell radius $\overline{R}$, extracellular $\overline{D}_e$ and intracellular $\overline{D}_i$ diffusion coefficients and intracellular volume fraction $\overline{\text{\textit{ICVF}}}$.

\subsubsection{Cell size and intracellular volume fraction estimation}
The impermeable Ball \& Sphere model accurately estimated the substrates parameters in the impermeable substrates, except for $D_i$. It overestimated and underestimated the cell size $R$ and the intracellular volume fraction $\text{\textit{ICVF}}$ as permeability increased, respectively  \citep{Afzali2021, Olesen2022}. This opposite evolution indicated how the models developed for impermeable tissues compensated for water exchange. Because the cell size limited the distance traveled by the particles in impermeable cells, the increase in diffusion distance due to permeability was compensated by either decreasing the proportion of the intracellular signal via a smaller $\overline{\text{\textit{ICVF}}}$ or increasing the maximal distance via a larger $\overline{R}$. This deterioration of the estimates was more significant for smaller cells, which is coherent with the observation that diffusion enters the exchange-dominated regime at a shorter diffusion time.

On the other hand, this effect was attenuated with the CEXI model, thanks to the exchange time capturing most of the exchange effect. At moderate permeability $(\kappa < 25 \frac{\mu m}{s})$, CEXI disentangled the effect of exchange and restriction from the DW-MRI signals, providing more stable estimates of $\overline{\text{\textit{ICVF}}}$ and $\overline{R}$ with the cell size and the permeability than the impermeable Ball \& Sphere model over this range of permeability. When diffusion was dominated by exchange, i.e. in the small cells $R<5 \mu m$, the mean cell size and the intracellular volume fraction were better estimated. When diffusion was not dominated by the exchange yet ($R > 5 \mu m)$, the $ADC(t)$ was strongly time-dependent, and the CEXI model assumptions were not valid. Consequently, the intracellular volume fraction estimates were less stable as a function of permeability.

\subsubsection{Sensitivity to the compartment diffusivity variations}
The compartment diffusivities were arguably the most difficult parameters to estimate due to the degeneracy of the solution \citep{Jelescu2016, Novikov2018} and the low sensitivity of the models to the intracellular diffusion coefficient \citep{Li2017}. In white matter, the intracellular diffusion coefficient was often considered faster than the extracellular diffusion coefficient \citep{Kunz2018, Dhital2019, Olesen2021}. Still, recent studies in gray matter suggested contradictory conclusions on which compartment had the fastest diffusivity ($D_e > D_i$ in ex-vivo \citep{Olesen2022} or $D_e < D_i$ in in-vivo \citep{Jelescu2022}). In our experiments, the impermeable model highlighted this degeneracy of the solution through unstable or estimates of the diffusivities. 

Conversely, the CEXI model showed sensitivity to $D_e$ and $D_i$ changes with a limited impact on the estimates $\overline{R}$ and $\overline{\text{\textit{ICVF}}}$, supporting that the contribution of each compartment to the total signal could be disentangled by including an exchange parameter in the model. At moderate permeability ($\kappa \leq 25 \mu m/s)$ and in the larger spheres ($R > 3 \mu m$), CEXI was sensitive to the variations of the intracellular diffusion coefficient. In the small spheres ($R < 3 \mu m$), however, the cells were too small for CEXI to be sensitive to $D_i$. Indeed, diffusion inside the intracellular compartment reached the Gaussian diffusion limit before the shortest diffusion time, irrespective of the true $D_i$.

\subsubsection{Permeability estimate}
In the previous section, we identified exchange-dominated and restriction-dominated regimes. We determined that the transition between those regimes occurred for the substrates of cell size around $4-5 \mu m$. In smaller cells, the exchange dominates; therefore, we should be sensitive to membrane permeability. The accurate estimation of the permeability with CEXI in this range of cell sizes confirmed this observation. At moderate permeability ($\kappa \leq 25 \mu m/s)$, CEXI estimated the permeability accurately with a small variance. The best estimates were obtained for substrates in the transition regime, i.e. cells of size $4-5 \mu m$. In larger cells, the estimates' variance was too large to be reliable.

\subsection{Recommendations}
In Sections~\ref{sec:discussion_signal} and~\ref{sec:discussion_model}, we showed the importance of probing tissue in the correct diffusion time frame. In permeable tissue, diffusion translates from a restriction-dominated to an exchange-dominated regime at a long enough diffusion time. To find a good range of diffusion times for the tissue under consideration, a preliminary time-dependency analysis of the apparent diffusion coefficient $ADC(t)$ and kurtosis $AKC(t)$ might be insightful. Indeed, we showed that diffusion moves from the restriction-dominated to the exchange-dominated regime at the peak value of the $AKC(t)$. Our CEXI model provided more robust estimates in this exchange-dominated regime than the impermeable Ball \& Sphere model in permeable substrates. In the largest cells or at high permeability values, estimation remained challenging even with CEXI. In the first case, diffusion was dominated by structure and not exchange. In the second case, the exchange was too fast. This finding suggested that future models should be developed by generalising the assumptions behind the K\"arger model to accurately estimate the microstructure parameters in highly permeable tissue. These observations show that the PGSE sequence is suitable for a specific exchange time. Alternative sequences, such as double-diffusion encoding \citep{Shemesh2012, Henriques2021} or stimulated echo sequence \citep{Karunanithy2019}, might be better adapted for tissue with a longer exchange time. On the other hand, oscillatory gradient spin echo sequence (OGSE) \citep{Reynaud2016} showed promising results in tissue with shorter exchange time.

\subsection{Towards lymph nodes imaging}
Based on previous conclusions and histology data, we could design a protocol for real DW-MRI data acquisitions on lymph nodes. The lymphocyte radius is around $3-5\mu m$, so the molecular diffusion in the lymphocytes should already be in the exchange regime at a diffusion time of about $30ms$. We could determine the lymphocyte permeability by comparing acquisitions to the simulations. In the example of malignant lymph nodes shown in Figure~\ref{fig:histology_substrate}, the cells are bigger with a radius around $6-10\mu m$. Hence, diffusion will be in the restriction-dominated regime for a diffusion time under $30ms$. We plan to acquire data with a longer diffusion time until we reach the exchange-dominated regime. We expect this threshold diffusion time to be correlated with cancerous tissue. Future work will combine membrane permeability and anisotropic diffusion to capture better the complexity of lymph node microstructure \citep{Ianus2020} and, consequently, determine if a model with more compartments is needed \citep{Stanisz1997}.

\section{Conclusion}\label{sec::conclusion}
This work showed, using simulations in numerical substrates of packed spheres, that the water exchange between the intracellular and extracellular spaces cannot be neglected in permeable tissues when diffusion is in the exchange-dominated regime. The time-dependency of the kurtosis and the signal could be used to identify the dominating diffusion regime and, ultimately, the most relevant biophysical model and the experimental parameters of the imaging protocol best suited to estimate its parameters. Additionally, the inherent bias in the estimates of the compartmentalized models for impermeable tissue was amplified in permeable tissue, even for very low permeability values. As an alternative, a two-compartment model of permeable tumors considering the water exchange between spherical cells and the extracellular space was evaluated, allowing us to simultaneously estimate the exchange time and cell size. Despite the improved performance compared to the impermeable model in the regime from low to moderate permeability levels, some limitations were found in highly permeable substrates, suggesting the need for a more general model of permeable tissue that accounts for the non-Gaussian diffusion in the compartments and the time-dependency of the diffusion coefficients.

\section*{Acknowledgments}
This work is supported by the Swiss National Science Foundation under grants 205320\_175974 and 205320\_204097. 

We acknowledge access to the facilities and expertise of the CIBM Center for Biomedical Imaging, a Swiss research center of excellence founded and supported by Lausanne University Hospital (CHUV), University of Lausanne (UNIL), Ecole Polytechnique Fédérale de Lausanne (EPFL), University of Geneva (UNIGE) and Geneva University Hospitals (HUG). Erick J. Canales-Rodríguez was supported by the Swiss National Science Foundation (Ambizione grant PZ00P2\_185814). 

\section*{Data Availability Statement}
The data used for this manuscript were simulated with an extension of the open-source Monte-Carlo simulator (\href{https://jonhrafe.github.io/MCDC_Simulator_public/}{https://jonhrafe.github.io/MCDC\_Simulator\_public/}) and can be shared upon request by emailing the corresponding author (remy.gardier@epfl.ch). 

\section*{Conflict of Interest Statement}
The authors declare that the research was conducted in the absence of any commercial or financial relationships that could be construed as a potential conflict of interest.

\section*{Author Contributions}
RG: Methodology, Coding, Simulations, Analysis, Writing, Visualization,
JLVH: Discussion about substrate generation and the choice of the parameters of the simulation, Writing - Review,
EJC-R: Methodology, Writing-Review and Editing,
IOJ: Discussion about the compartmentalized model and mesoscopic disorder theory, Writing-Review and Editing,
GG: Methodology, Experimental design, Writing-Review and Editing,
JR-P: Methodology, Experimental design, Discussion about simulations, Active contribution to the analysis of the results, Supervision, Writing - Review and Editing,
J-PT: Supervision, Funding, Writing-Review 

\begin{table*}[t]%
     \caption{\textbf{Parameters of the Monte-Carlo diffusion simulations}. The experiments are detailed in Section \ref{sec:method_method}. The voxel side length and the intracellular volume fraction (ICVF) are the same for all substrates. $R_s$ is the desired mean cell radius of the sphere populations, $R_{mean}$ is the effective mean cell radius, $R_{small} = \left(\frac{< R^7 > }{<R^3>}\right)^{\frac{1}{4}}$ and $R_{NP} = \left(\frac{< R^5 > }{<R^3>}\right)^{\frac{1}{2}}$.}\label{tab:experiments}   
     
     \begin{tabular*}{\textwidth}{@{\extracolsep\fill}lccccc@{\extracolsep\fill}}
        \toprule\multicolumn{6}{c@{}}{\textbf{Substrate parameters}}\\
        \midrule& $S_1$ & $S_2$  & $S_3$  & $S_4$ & $S_5$\\
        \midrule Voxel side length ($\mu m$)  & \multicolumn{5}{@{}c@{}}{$100$}\\
        ICVF           & \multicolumn{5}{@{}c@{}}{$0.65$} \\
        $R_s$ ($\mu m$)  & 2     & 3     & 4     & 5 & 8 \\
        $R_{mean}$ ($\mu m$)    & 1.9   & 3.0   & 4.0   & 5.0 & 8.1  \\
        $R_{small}$ ($\mu m$)    & 2.2   & 3.1   & 4.1   & 5.1 & 8.3 \\
        $R_{NP}$ ($\mu m$)    & 2.1   & 3.1   & 4.0   & 5.1 & 8.2\\
        \midrule\multicolumn{6}{c@{}}{\textbf{Simulation parameters}}\\
        \midrule $D_{e,0} (\frac{\mu m^2}{ms})$ &  $D_{i,0} (\frac{\mu m^2}{ms})$ &  $\kappa (\frac{\mu m}{s})$ &  $\rho (\frac{part}{\mu m^3})$ & $\delta t (\mu s)$\\ 
         1 ,2     & 1, 2  & 0, 10, 25, 50 & 0.5   & 5 \\
        \midrule\multicolumn{6}{c@{}}{\textbf{PGSE sequence parameters}}\\
        \midrule $\Delta (ms)$       &  $\delta (ms)$    &  $TE (ms)$    &  $b (\frac{ms}{\mu m^2})$     & N\\
        12, 20, 30, 40      & 4.5               & 50            & 1,   2.5, 4, 5.5, 7           & 24\\
        \bottomrule
    \end{tabular*}
\end{table*}

\begin{table*}[t]%
    \caption{\textbf{Results of the mesoscopic disorder fit.} Parameters of the curves shown in Fig.\ref{fig:ADC_AKC_disorder}. The cell size $R$ is in $\mu m$, the extracellular diffusion coefficient $D_{e, 0}$ and the parameters $ADC_{\infty}$, $\text{\textit{MSE}}_D$ are in $\mu m^2/ms$, the parameter $A_D$ is in $\mu m^2$ and the parameter $A_K$ is in $ms$.}\label{tab:mesoscopic_fit}    
    \begin{tabular*}{\textwidth}{@{\extracolsep\fill}lccccccc@{\extracolsep\fill}}
        \toprule\multicolumn{2}{@{}c@{}}{\textbf{Simulations}} & \multicolumn{3}{@{}c@{}}{$ADC_{ex} = ADC_{\infty} + A_D /t$}& \multicolumn{3}{@{}c@{}}{$AKC_{ex} = AKC_{\infty} + A_K /t $}\\ 
        $R$ & $D_{e, 0}$ & $A_D$ & $ADC_{\infty}$ & $\text{\textit{MSE}}_D$ &  $A_K$ & $AKC_{\infty}$ &$\text{\textit{MSE}}_K$\\\cmidrule{1-2} \cmidrule{3-5} \cmidrule{6-8} 
        2 & 1 & $0.00012$ & $0.67$ & 0.0002 & $1.25$ & $0.003$  & 0.0012  \\
        2 & 2 & $0.00012$ & $1.33$ & 0.0004 & $0.49$ & $0.008$  & 0.0001\\
        4 & 1 & $0.00027$ & $0.67$ & 0.0005 & $3.84$ & $0.013$   & 0.0002 \\
        4 & 2 & $0.00011$ & $1.34$ & 0.0007 & $0.72$ & $0.005$  & 0.0020\\
        8 & 1 & $0.00107$ & $0.68$ & 0.0003 & $5.21$ & $0.113$  & 0.0040\\
        8 & 2 & $0.00144$ & $1.35$ & 0.0004 & $6.17$ & $0.029$  & 0.0027\\\bottomrule
        \end{tabular*}
\end{table*}

\section*{Supporting figures}
\paragraph*{\textbf{Figure S1: BALL\&Sphere impermeable - Signal fitting.}} The plots show the fitting of the BALL\&Sphere model to the simulated signals with the diffusion time for different $b$ (symbols), permeability (color) and cell size (columns).

\paragraph*{\textbf{Figure S2: CEXI - Signal fitting.}} The plots show the fitting of the CEXI model to the simulated signals with the diffusion time for different $b$ (symbols), permeability (color) and cell size (columns).

\paragraph*{\textbf{Figure S3: Model estimates.}} The plots show the estimates of the Ball \& Sphere (A-D) and CEXI (E-H) models from 30 DW-MRI signals with a \textit{SNR} of 30.

\paragraph*{\textbf{Figure S4: BIC of model estimates.}} BIC of the impermeable Ball \& Sphere model and CEXI for with increasing permeability, at different SNR.

\paragraph*{\textbf{Figure S5:Ball \& Sphere estimates - Low ICVF - Noiseless.}} The plots show the estimates of the Ball \& Sphere model in substrates with an ICVF of 0.2 from on noiseless signals.

\paragraph*{\textbf{Figure S6: Ball \& Sphere estimates - Low ICVF - Noisy.}} The plots show the estimates of the Ball \& Sphere model in substrates with an ICVF of 0.2 from 30 DW-MRI signals with a \textit{SNR} of 30.

\paragraph*{\textbf{Figure S7: CEXI estimates - Low ICVF - Noiseless.}} The plots show the estimates of the CEXI model in substrates with an ICVF of 0.2 from on noiseless signals.

\paragraph*{\textbf{Figure S8: CEXI estimates - Low ICVF - Noisy.}} The plots show the estimates of the CEXI model in substrates with an ICVF of 0.2 from 30 DW-MRI signals with a \textit{SNR} of 30.

\clearpage
\bibliographystyle{preprint} 
\bibliography{biblio}
\vfill\pagebreak

\end{document}

%% file: newcommands.tex
\usepackage{booktabs}%

\def\keyFont{\fontsize{8}{11}\helveticabold }